\documentclass[10pt]{article}

\pdfoutput=1

\usepackage{color}

\usepackage{thumbpdf}
\usepackage{jheppub}
\usepackage{esint,shuffle,psfrag}
 \usepackage[utf8]{inputenc}

\allowdisplaybreaks
\usepackage{varioref}
\usepackage{amsmath,amsfonts,amsthm,mathrsfs}
\usepackage{enumerate}
\usepackage{fancyvrb}
\usepackage{verbatim}
\usepackage{wrapfig}
\usepackage{appendix}
\usepackage{amstext}
\usepackage{amssymb}
\usepackage{graphicx}
\usepackage{color}
\usepackage{varioref}
\usepackage{multirow,graphics}
\usepackage{epstopdf}
\usepackage{bbm}
\usepackage{slashed}
\usepackage{psfrag}
\usepackage{relsize}
\usepackage{bbm}
\usepackage{mathrsfs}
\usepackage{epsfig}
\usepackage{upgreek}

\usepackage{mathtools}

       \usepackage{dutchcal}
       \usepackage{dsfont}

\DeclareMathAlphabet{\mathpzc}{OT1}{pzc}{m}{it}

\usepackage{tikz}
\usepackage{subcaption}

\usepackage{mathtools}
\usepackage{bigints}

\usetikzlibrary{decorations}

\usepackage{amstext}
\usepackage{amssymb}
\usepackage{amsmath}
\usepackage{verbatim}

\newcommand{\alg}[1]{\mathfrak{#1}}

\newcommand{\bunderline}[1]{\underline{#1\mkern-1mu}\mkern5mu }

\def\mr{{\bunderline{r}}}

\newcommand{\bea}{\begin{eqnarray}}
\newcommand{\eea}{\end{eqnarray}}

\def \nn {N}
\def\la{\label}

\def\qq{{\mathcal q}}
\def\gQ{{\mathpzc{Q}\hspace{.07em}}}
\def\mI{\mathbbm{1}}

\def\a {\alpha}
\def\b {\beta}

\newcommand{\sfrac}[2]{{\textstyle\frac{#1}{#2}}}

\numberwithin{equation}{section}

\usepackage{tikz}
\usetikzlibrary{plotmarks,calc,decorations, decorations.pathmorphing, patterns}
\usetikzlibrary{arrows}
\tikzstyle dynkin node=[very thick,shape=circle,draw,inner sep=0pt,minimum size=5mm]
\tikzstyle dynkin line=[very thick]
\tikzstyle inverse line=[gray,line width=1.46pt,line cap=round, dash pattern=on 0pt off 2\pgflinewidth]
\tikzstyle red phase=[red,decoration={snake,amplitude=0.1mm,segment length=1.6mm},decorate]
\tikzstyle blue phase=[blue,decoration={snake,amplitude=0.1mm,segment length=0.9mm},decorate]
\tikzstyle green phase=[green,decoration={snake,amplitude=0.1mm,segment length=0.9mm},decorate]
\tikzstyle brown phase=[brown,decoration={snake,amplitude=0.1mm,segment length=0.9mm},decorate]
\newcommand{\boundellipse}[3]
{(#1) ellipse (#2 and #3)
}
\usetikzlibrary{decorations.pathmorphing}
\tikzstyle arrow=[thick,rounded corners=18pt,-latex]
\tikzstyle box=[draw,rounded corners,outer sep=4pt]
\tikzstyle B node=[outer sep=0pt]
\tikzstyle Q node=[inner sep=1pt,outer sep=0pt]
\definecolor{purple_nice}{rgb}{0.4,0.2,0.7}
\definecolor{fuel_blue}{RGB}{42,162,185}

\def\<{\langle}
\def\>{\rangle}

\newcommand{\rr}{\mathcal{r}}

\makeatletter
\@ifundefined{usebibtex}{} {}
\makeatother

\def\Jd{{\alg g}^*}
\def\J{{\alg g}}

\def\double{{\mathscr D}}

\newcommand{\ia}{\mathbf{a}}
\newcommand{\ic}{\mathbf{c}}

\def\M{\mathcal{M}}
\def\N{\mathcal{N}}
\def\P{\mathscr{M}}
\def\var{{\varkappa}}

\def\rC{{\raisebox{1pt}{\scriptsize rec}}}
\def\sC{{\raisebox{1pt}{\scriptsize s}}}
\def\el{\ell}
\def\S{{\rm S}}
\def\W{\mathscr{W}}
\def\t{\uptau}

\title{\Large Hyperbolic spin Ruijsenaars-Schneider model from \\Poisson reduction$^*$}
\note[$*$]{Invited contribution to the special issue of the ``Proceedings of the Steklov Institute of Mathematics" dedicated to the 80th anniversary of Prof. Andrei Slavnov.}

\author[a]{Gleb Arutyunov}
\author[a]{and\,~Enrico Olivucci}

\affiliation[a]{II. Institut f\"ur Theoretische Physik, Universit\"at Hamburg, Luruper Chaussee 149, 22761
Hamburg, Germany\\
Zentrum  f\"{u}r  Mathematische  Physik,  Universit\"{a}t  Hamburg,  Bundesstrasse  55,  20146  Hamburg,Germany}

\emailAdd{gleb.arutyunov@desy.de}  
\emailAdd{enrico.olivucci@desy.de}

\abstract{
We derive a Hamiltonian structure for the $N$-particle hyperbolic spin Ruijsenaars-Schneider model by means of Poisson reduction of a suitable initial phase space. This phase space is realised as the direct product of the Heisenberg double of a factorisable Lie group with another symplectic manifold 
that is a certain deformation of the standard canonical relations for $N\ell$ conjugate pairs of dynamical variables. We show that the model enjoys 
the Poisson-Lie symmetry of the spin group ${\rm GL}_{\ell}({\mathbb C})$ which explains its superintegrability. Our results are obtained in the formalism
of the classical $r$-matrix and they are compatible with the recent findings on the different Hamiltonian structure of the 
model established in the framework of the quasi-Hamiltonian reduction applied to a quasi-Poisson manifold.
}

\begin{document}

\begin{flushright}\small{ZMP-HH-19-9}\end{flushright}

\maketitle

\flushbottom

\section{Introduction}
The Ruijsenaars-Schneider (RS) integrable models \cite{Ruijsenaars:1986vq,Ruijsenaars:1986pp} continue to deliver 
rich mathematical structures that are worth further exploring.  One particular aspect concerns the introduction of spin
degrees of freedom. Recall that a spin generalisation of the RS model with the most general elliptic potential 
was proposed in \cite{Krichever:1995zw} as a dynamical system describing the evolution of poles of elliptic solutions 
of the non-abelian 2d Toda chain. This is a system of $N$ particles on a line with internal degrees of freedom represented 
by two $\ell$-dimensional vectors attached to each of the particles. The proposed spin RS model is given in terms of 
equations of motion for the particle coordinates $q_i$, $i=1,\ldots, N$ and the spin variables\footnote{We follow the notation of \cite{Arutyunov:1997ey}.} $\mathbf{a}_{i\a}$ and $\mathbf{c}_{\a i}$, where $\a=1,\ldots, \ell$.
The knowledge of the equations of motion contains but unfortunately does not immediately yield the Hamiltonian structure
behind this dynamical system.

In \cite{Arutyunov:1997ey} we established the underlying  Hamiltonian structure 
for the case of rational degeneration of the elliptic spin RS model. This was done by relaying on the observation that goes back to 
\cite{KKS} and further developed in \cite{Gorsky:1993dq}-\cite{Feher:2018pmu} that the Calogero-Moser-Sutherland and Ruijsenaars-Schneider models can be obtained by means of the Hamiltonian or Poisson reduction procedure applied to a suitably chosen initial phase space. In the case of the rational spin RS model 
the suitable initial phase space $\P$ appears to be the direct product $\P=T^*G\times \Sigma $, where $T^*G$ is the cotangent bundle to a Lie group $G$ with the Lie algebra $\J$ and $\Sigma$ is the symplectic manifold of $N\ell$ pairs of canonical variables (oscillators). This phase space is a Poisson manifold which carries the Hamiltonian action of $G$. Choosing $G={\rm GL}_N({\mathbb C})$ the Hamiltonian reduction of $\P$ by the action of $G$ yields the desired Poisson structure of the spin RS model 
\cite{Arutyunov:1997ey}. The Poisson brackets of the invariant spin variables appear rather involved. Although it was possible to guess a natural generalisation of the Poisson structure for 
``collective" spin variables $f_{ij}=\sum_{\a}\mathbf{a}_{i\a}\mathbf{c}_{\a j}$
to the hyperbolic spin RS model, the progress of finding the Poisson structure of individual spins in the hyperbolic case was delayed for years.
Quite recently this structure has been found \cite{Chalykh:2018wce} confirming the conjecture in \cite{Arutyunov:1997ey} on the brackets of collective spin variables.
The approach of \cite{Chalykh:2018wce}, see also \cite{Fairon:2018zgn,Chalykh:2017urw}, is based on the quasi-Hamiltonian reduction procedure, where one starts from 
an initial  manifold $\P$ supplied with a quasi-Poisson structure and which carries a free action of a Lie group $G$. 
Although $\P$ is not Poisson, the quotient $\P/G$ inherits the well-defined Poisson structure from the quasi-Poisson structure on $\P$.    
Picking as $\P$ a representation space of a framed Jordan quiver, it was shown in \cite{Chalykh:2018wce}  that the 
reduction of this by $G$ yields the Poisson structure of invariant spins that perfectly fits the hyperbolic (trigonometric complex)
spin RS model. The Liouville integrability and superintegrability (degenerate integrability) of the spin RS model also follow from this approach.

Having established these nice results, one still may wonder if there would exist a conventional way of getting the spin hyperbolic RS model
by the usual Poisson reduction but applied to a more complicated initial phase space being the next in the deformation hierarchy after $T^*G\times \Sigma$
responsible for the rational model. Indeed, the spinless hyperbolic RS model follows from the Poisson reduction applied to the Heisenberg double $D_+(G)$ of $G$,
as has been recently discussed in \cite{Arutyunov:2019wuv}.
The Poisson structure of the Heisenberg double \cite{SemenovTianShansky:1985my} is a deformation of the one of $T^*G$.  From the point of view of the deformation theory,
it is then natural to replace the moment map on $\Sigma$, taking values into the dual Lie algebra $\J^*$, with a non-abelian moment map defined on a suitable deformation of $\Sigma$ and which takes values in the dual Poisson-Lie group $G^*$. The main question is how to realise the quadratic Poisson structure of $G^*$
in terms of $N\ell$-pairs of oscillators that should replace those used to represent the linear  Kirillov-Kostant bracket in the rational case. In this paper we solve this problem 
and reconstruct the spin hyperbolic RS model in the standard framework of the Poisson reduction.

The main tool in our approach is a Poisson pencil of a constant and quadratic Poisson structures on an {\it oscillator manifold} $\Sigma_{N,\ell}$ 
spanned by $2N\ell$ dynamical variables  $a_{i\a},b_{\a i}$.
When the coefficient $\var$ in front of the quadratic structure vanishes, one obtains the standard canonical relations of the $N\ell$ conjugate pairs. 
In fact there are two different quadratic structures, to distinguish between them we label the corresponding Poisson manifolds as $\Sigma^{\pm}_{N,\ell}$.
These Poisson manifolds carry Poisson actions of two different Poisson-Lie groups -- the {\it particle group} ${\rm GL}_N(\mathbb{C})$ and the {\it spin group} ${\rm GL}_{\ell}(\mathbb{C})$, acting by linear transformations on the oscillator indices $i$ and $\alpha$, respectively.
Starting from the initial phase space $\P=D_+(G)\times \Sigma_{N,\ell}^{\pm}$ and reducing this manifold by the action of the particle group, we obtain the 
spin RS model with the Poisson structure inherited from that on $\P$. The equations of motion for the spins are the same regardless of which manifold $\Sigma^{\pm}_{N,\ell}$ we use, and they coincide with those that follow from the Poisson structure of spins found in \cite{Chalykh:2018wce} through the quasi-Hamiltonian reduction. The construction of conserved quantities, both Poisson commutative and non-commutative, is straightforward and follows 
the same pattern as in the rational case.
The spin group continues to act on the reduced phase space as a Poisson-Lie symmetry and its presence explains the superintegrability of the model. 
In fact, there are higher symmetries  whose generators are  polynomial in the spin variables and which arise from conjunction of the spin symmetries 
with abelian symmetries generated by higher commuting charges. We show that the Poisson structure of the currents
encoding these symmetries is a quadratic deformation of the linear bracket of the rational model. This quadratic part appears as an affine 
version of the Poisson-Lie structure on $G^*$.
 
Concluding the brief discussion of our approach, we point out that it would be interesting to extend it to account for the most general elliptic spin model.   Also,
since we are building on the classical $r$-matrix formalism, the recognition of various $r$-matrix structures 
might help  to pave the way for quantising the spin model which currently remains another open problem.
   
The paper is organised as follows. In the next section we recall the necessary facts about the Heisenberg double.   
In section 3 we introduce the oscillator manifold. 
 In section 4 we discuss the Poisson action of a Poisson-Lie group on the product of two manifolds. 
 In section 5 
 we solve the 
 moment map equation obtaining the Lax matrix of the spin RS model on the reduced phase space. 
 The Poisson brackets of $G$-invariant variables are studied in section 6 and section 7
 is devoted to the discussion of symmetries of the model responsible for its superintegrable status. We conclude this section by showing what superintegrability 
 implies for solvability of the equations of motion. Some technical details are collected in appendix. 
 All the considerations in the paper are done in the context of holomorphic integrability.

\section{Heisenberg Double}
We start with recalling the construction of the double of a factorisable Lie bialgebra.
Let $G$ be a Lie group with the Lie algebra $\J$. Denote by $\Jd$ the dual of $\J$. We assume that $(\J,\Jd)$ is a factorisable Lie bialgebra and we use the 
corresponding invariant form on $\J$ to identify $\Jd\simeq \J$. The double $\double$ of $(\J,\Jd)$ can be identified with $\double=\J\oplus \J$ supplied with the Lie algebra 
structure of the direct sum of two copies of the Lie algebra. The Lie algebra $\J\subset \double$ is embedded in $\double$ as the diagonal sublagebra, while the Lie subalgebra $\J^*$ 
is identified inside $\double$ as a subset
\bea
\nonumber
(X_+,X_-)=(\hat{\rr}_+X,\hat{\rr}_-X)\subset \double \, , ~~~\forall X\in \Jd\simeq\J\, . 
\eea
Here $\hat{\rr}_{\pm}=\hat{\rr}\pm \sfrac{1}{2}\mI$ are two linear operators, $\hat{\rr}_{\pm}:\,\Jd\to\J_{\pm}\subset \J$, constructed from a skew-symmetric split solution $\rr\in \J\,\wedge\,\J$ of the modified Yang-Baxter equation.
Any $X\in \J$ has a unique decomposition $X=X_+-X_-$.

Let $D= G\times G$ be the double Lie group corresponding to $\double$. The connected Lie group $G^*$ corresponding to the Lie algebra $\J^*$
is embedded in $D$ as $G^*\simeq (u_+,u_-)\subset D$ by extending the Lie algebra homomorphisms given by $\hat{\rr}_{\pm}$. Here $u_{\pm}\in G_{\pm}$, where 
$G_{\pm}$ are the corresponding subgroups of $G$. In the following we assume the existence of a global diffeomorphism $\sigma:\, G^*\simeq G$, 
\bea
\label{factor}
\sigma(u_+,u_-)=u_+u_-^{-1}=u\, ,
\eea
such that the factorisation problem (\ref{factor}) has a unique solution for any $u\in G$.

Now we introduce the Heisenberg double $D_+(G)$ of $G$.
Consider a pair of matrices $(A,B)\in D$, $A,B\in G$.
The entries of $A,B$ can be regarded as generators of the coordinate ring of the algebra or regular functions 
on $D$.  The Heisenberg double $D_+(G)$
is $D$ viewed as a Poisson manifold with the following Poisson relations between the generators 
\bea
\label{PBAB}
\begin{aligned}
\frac{1}{\var}\{A_1,A_2\}\,&=\, -\rr_- \,A_1 A_2- A_1 A_2 \,\rr_+ + A_1 \,\rr_- \, A_2+ A_2 \,\rr_+\, A_1 \, ,\\
\frac{1}{\var}\{B_1,B_2\}\,&=\, -\rr_-\, B_1 B_2- B_1 B_2\, \rr_+ +B_1\, \rr_-\, B_2+ B_2 \,\rr_+\, B_1\, , \\
\frac{1}{\var}\{A_1,B_2\}\,&=\, -\rr_- \,A_1 B_2-A_1 B_2\, \rr_- + A_1 \,\rr_- \, B_2+B_2 \,\rr_+ \,A_1\, , \\
\frac{1}{\var}\{B_1,A_2\}\,&=\, -\rr_+\, B_1 A_2- B_1 A_2 \, \rr_+ + B_1\, \rr_-\, A_2+A_2 \,\rr_+ \,B_1\, .
\end{aligned}
\eea
where $\var$ is a complex parameter.
Here $\rr_{\pm}$ are two canonical solutions of the classical Yang-Baxter equation associated with the factorisable Lie algebra $\J$; they correspond to the operators 
$\hat{\rr}_{\pm}$. 

In this work we are primarily interested in the case $G={\rm GL}_{\nn}({\mathbb C})$ for which the matrices $\rr_{\pm}$ are  
\begin{align}
\la{rr-mat}
\rr_\pm = \pm\frac{1}{2} \sum_{i=1}^N E_{ii} \otimes E_{ii} \pm \sum_{i\lessgtr j}^N E_{ij} \otimes E_{ji}\, .
\end{align}
Here $E_{ij}$ are the standard matrix unities, $(E_{ij})_{kl}=\delta_{ik}\delta_{jl}\, .$
We also recall that
\begin{align}
\rr_{\pm \, 21}\,=\,-\rr_{\mp\, 12}\, , \qquad \rr_+-\rr_- = C_{12} = \sum_{i,j=1}^N E_{ij}\otimes E_{ji}\,,
\end{align} and introduce \(\rr=\sfrac{1}{2}(\rr_+ +\rr_-)\),  which is a skew-symmetric split solution to the modified classical Yang-Baxter equation mentioned above. 

The Heisenberg double (\ref{PBAB}) carries a Poisson action of a Poisson-Lie group $G$ 
\bea
\la{adj_act}
A\to hAh^{-1}\, , ~~~~B\to hBh^{-1}\, ,~~~h\in G\, .
\eea
The Poisson-Lie structure of $G$ is given in terms of the Sklyanin bracket
\bea
\la{PL_G}
\{h_1,h_2\}=-\var\, [\rr_{\pm},h_1h_2]\, ,~~~h\in G\, .
\eea 
The non-abelian moment map for this action   $(\M_+,\M_-)$ takes values in the group $G^*$. Under $\sigma$ it maps onto an element $\M=\M_+\M_-^{-1}\in G$, where  
\bea
\la{mmHD}
\M=BA^{-1}B^{-1}A\, .
\eea
The Poisson algebra between the entries of $\M$ is
\bea
\la{M_double_alg}
\frac{1}{\var}\{\M_1,\M_2\}=-\rr_+ \M_1\M_2-\M_1\M_2\rr_- + \M_1\rr_-\M_2 + \M_2\rr_+\M_1\, .
\eea

The Poisson algebra (\ref{PBAB}) has two obvious involutive subalgebras - one is generated by 
${\rm Tr} A^k$ and the other by ${\rm Tr}B^k$, where $k\in \mathbb{Z}$. There is yet another involutive family which plays an essential role in this work, 
namely, 
\bea
\la{ham_before}
H_k={\rm Tr}(BA^{-1})^k={\rm Tr}(A^{-1}B)^k\, , ~~~k\in \mathbb{Z}\, .
\eea
The fact that $\{H_k,H_m\}=0$ for any $k,m\in \mathbb{Z}$ can be verified by direct computation. A deeper observation is that the map
\bea
A\to A\, , ~~~B\to BA^{-1}\, ,
\eea
is a canonical transformation, {\it i.e.} under this map the Poisson structure (\ref{PBAB}) remains invariant. Note that all the involutive families mentioned above are 
generated by invariants of the adjoint action (\ref{adj_act}).


In the following we need two facts about the group $G^*$. First, $G^*$ is a Poisson-Lie group. In terms of the generators $u_{\pm}\in G_{\pm}\subset G$ the corresponding Poisson-Lie structure is given by the following Poisson brackets  
\bea
\la{PL_Gstar}
\begin{aligned}
&\frac{1}{\var}\{u_{\pm 1},u_{\pm 2}\}=-[\rr, u_{\pm 1}u_{\pm 2}]\, , ~~~   & \frac{1}{\var}\{u_{\pm 1},u_{\mp 2}\}=-[\rr_\pm ,u_{\pm 1}u_{\mp 2}]\, .
\end{aligned}
\eea
Under the map (\ref{factor}), these brackets endow $G$ with the structure of a Poisson manifold given by the Semenov-Tian-Shansky bracket \cite{SemenovTianShansky:1985my}
\bea
\la{PL_as_G}
\frac{1}{\var}\{u_1,u_2\}=-\rr_+u_1u_2- u_1u_2\rr_-+u_1\rr_-u_2+u_2\rr_+u_1\, .
\eea   
Comparing (\ref{M_double_alg}) with (\ref{PL_as_G}) shows that the Poisson algebra of $\M$ is given by the Semenov-Tian-Shansky bracket.

The product in $G^*$ induces under (\ref{factor}) a new product in $G$ which we denote by $\star$. For any $u,v\in G^*$ it is defined as 
\bea
\la{star_prod}
v\star u=v_+u_+u_-^{-1}v_-^{-1}=v_+uv_-^{-1}\, .
\eea
where $u_{\pm}$ and $v_{\pm}$ are solutions of the factorisation problems $u=u_+u_-^{-1}$ and $v=v_+v_-^{-1}$. The Poisson-Lie structure of $G^*$ is then 
encoded in the following relation 
\bea
\nonumber
\{v_1\star u_1, v_2\star u_2\}=\{v_{+1}u_1v_{-1}^{-1},v_{+2}u_2v_{-2}^{-1}\}
=\{u_1,u_2\}
(v\star u)\, ,
\eea
where the bracket of $u$'s is  (\ref{PL_as_G}), while the brackets of $v_{\pm}$ are evaluated according to (\ref{PL_Gstar}).

Second, the Poisson-Lie group $G$ acts on $G^*$ by dressing transformations \cite{SemenovTianShansky:1985my}. Modelling $G^*$ over $G$, these transformations take the form of the adjoint action\footnote{This is in fact the coadjoint action of $G$ on $G^*$.}
\bea
\la{dress_tr}
u\to huh^{-1}\, , ~~~~h\in G\, ,
\eea 
and they are Poisson maps of the Semenov-Tian-Shansky bracket provided the Poisson-Lie structure on $G$ is given by  (\ref{PL_G}).
The non-abelian moment map of this action is $u$. It is well known that the symplectic leaves of (\ref{PL_as_G}) coincide with the 
orbits of  (\ref{dress_tr}).

\section{Oscillator manifold}
As the next step, we introduce a manifold $\Sigma_{N,\el}$ as the product of two linear spaces of all rectangular $N\times \el$-matrices 
\begin{align}
\label{spinman}
\Sigma_{N,\el} \,=\, {\rm Mat}_{N,\el}(\mathbb{C})\times {\rm Mat}_{\el,N}(\mathbb{C})\, ,
\end{align}
where \(N\) is the number of particles of the model and \(\el\) is the length of spin vectors. Let \((a,b)\) be two arbitrary $N\times \el$- and $\el\times N$-matrices.
Their entries 
\begin{align}
a_{i\alpha} \equiv (a)_{i\alpha} \, ,\,\, \, b_{\alpha j} \equiv (b)_{\alpha j}\, \qquad i=1,\dots,N\,,\quad \alpha=1,\dots,\el\,.
\end{align}
provide a global coordinate system on $\Sigma_{N,\el}$. We call $a_{i\alpha}$ and $b_{\alpha j}$ oscillators and refer to $\Sigma_{N,\el}$
as to an oscillator manifold. 

Now we endow $\Sigma_{N,\el}$ with two different $\pm$-structures  of a Poisson manifold $\Sigma_{N,\el}^{\pm}$ by defining the following Poisson brackets $\{\, ,\, \}_{\pm}$ between oscillators
\bea
\begin{aligned}
\label{PBoscill}
\{a_1,a_2\}_{\pm}\,&=\, \var\left(\,\rr \, a_1 a_2 \mp   \,a_1 a_2 \,\rho\right) \, , \\
\{b_1,b_2\}_{\pm}\,&=\, \var\left( \, b_1 b_2 \, \rr \, \mp\, \rho \, b_1 b_2 \right) \, , \\
\{a_1,b_2\}_{\pm}\,&=\, \var\left(-b_2\, \rr_+ \, a_1 \, \pm \, a_1 \,\rho_{\mp}  \,b_2 \, \right)\,-   C_{12}^\rC \, , \\
\{b_1,a_2\}_{\pm}\,&=\, \var\left(-b_1\, \rr_- \, a_2 \, \pm \, a_2  \,\rho_{\pm}\, b_1 \,\right)\, +  C_{21}^\rC\, .
\end{aligned}
\eea
Here we have introduced a ``rectangular split Casimir"
\begin{align}
C_{12}^\rC= \sum_{i=1}^{N}\sum_{\a=1}^\el E_{i \alpha} \otimes E_{\alpha i}\, ,\qquad \, 
\end{align}
where $(E_{i \alpha})_{j\beta} = \delta_{ij} \delta_{\alpha\beta}$. 
The matrices $\rho_{\pm}$ are the following analogues of $\rr_{\pm}$ in the spin space 
\begin{align}
\la{rho_m}
\rho_\pm = \pm\frac{1}{2} \sum_{\a=1}^\el E_{\a\a} \otimes E_{\a\a} \pm \sum_{\a\lessgtr \b}^\el E_{\a\b} \otimes E_{\b\a}\, 
\end{align}
and $\rho=\sfrac{1}{2}(\rho_+ +\rho_-)$. One also has 
\bea
\rho_+-\rho_- = C_{12}^\sC = \sum_{\a,\b=1}^\el E_{\a\b}\otimes E_{\b\a}\, .
\eea

For $\var=0$ the brackets (\ref{PBoscill}) turn into the standard oscillator algebra formed by $N\ell$ pairs of canonically conjugate variables
\bea
\la{canon}
\{a_{i\a},b_{\beta j} \}=-\delta_{ij}\delta_{\a\beta}\, .
\eea
The brackets (\ref{PBoscill}) satisfy the Jacobi identity for any $\varkappa$, {\it i.e.} the constant and quadratic structures in (\ref{PBoscill})  form a Poisson pencil being a one-parametric deformation of the canonical
relations (\ref{canon}). 
It remains to note that if we define 
\bea
\la{om_osc}
\omega=\mI+\var ab\, ,
\eea
where $ab$ is an $N\times N$-matrix being a natural product of two rectangular matrices, then due to (\ref{PBoscill}), $\omega$ will satisfy 
the Poisson algebra 
\bea
\la{PL_as_G_om}
\frac{1}{\var}\{\omega_1,\omega_2\}=\rr_+\omega_1\omega_2 +\omega_1\omega_2\rr_--\omega_1\rr_-\omega_2-\omega_2\rr_+\omega_1\, ,
\eea   
which is different from (\ref{PL_as_G}) by an overall sign only. 
In particular, the contribution of the spin matrices $\rho, \rho_{\pm}$ completely decouples. 
Thus, formulae (\ref{om_osc}) give a realisation of the Semenov-Tian-Shansky bracket in terms of the oscillator algebra (\ref{PBoscill}). We also point out the 
Poisson relations between $\omega$ and oscillators
\bea
\la{omega_os}
\begin{aligned}
\frac{1}{\var}\{\omega_1,a_2\}&=(\rr_+\omega_1-\omega_1\rr_-)a_2 \, , ~~~~~
\frac{1}{\var}\{\omega_1,b_2\}=-b_2(\rr_+\omega_1-\omega_1\rr_-) \,  .
\end{aligned}
\eea
In deriving (\ref{PL_as_G_om}) and (\ref{omega_os}) one has to use the relations
\bea
\nonumber
a_1C_{21}^\rC=C_{12}a_2\, , \qquad  C_{12}^\rC b_1=b_2C_{12}\, , \qquad C_{12}^\sC b_1b_2=b_1b_2C_{12}\, .
\eea

Importantly,  one can now verify that if we allow $G$ to act infinitesimally on oscillators as 
\begin{align}
\label{act_ab}
\delta_X a_{i\alpha}\,=\, ({\rm Ad}^*_{\omega} X\, a)_{i\a} \qquad \delta_X b_{\alpha i}\,=\, -( b \, {\rm Ad}^*_{\omega} X\,)_{\a i}\, , \qquad X\in \J\,,
\end{align}
then this action $G\times \Sigma_{N,\el}^{\pm}\to \Sigma_{N,\el}^{\pm}$ is a mapping of Poisson manifolds provided $G$ is equipped with the Sklyanin bracket \eqref{PL_G}.
Here ${\rm Ad}^*_{g}X$ for $g\equiv (g_+,g_-)\in G^*$ is the coadjoint (dressing) action of $G^*$ on the Lie algebra $\J$.
If we factorise $\omega = \omega_+\omega_-^{-1}$ according to \eqref{factor}, then $(\omega_+^{-1},\omega_-^{-1})\in G^*$ 
is the moment map for the Poisson action (\ref{act_ab}). Under (\ref{factor}) it defines the following element of $G$
\begin{align}
\label{MMosc}
\N\,=\, \omega_+^{-1}\omega_- \,\in\,G\, .
\end{align}  
The fact that $\N$ generates the action (\ref{act_ab}) can be deduced from the Poisson brackets (\ref{omega_os})
together with the fact that $\omega\star\{\N,\,.\,\}=-\{\omega,\,.\,\}\star \N$. The Poisson algebra of $\N$ coincides with (\ref{PL_as_G}).
\smallskip

Further, the oscillator manifolds carries an action of the spin Poisson-Lie group $S={\rm GL}_{\ell}(\mathbb{C})$
\bea
\la{int_group}
a_{i\alpha}\,\longrightarrow\, (ag)_{i\alpha}\, , \qquad b_{\alpha i}\,\longrightarrow\, (g^{-1}b)_{\alpha i}\, , ~~~g\in S\, .
\eea
This action is Poisson provided the Poisson-Lie structure on $S$ is taken for $\Sigma_{N,\el}^{\pm}$ to be 
\bea
\la{spinS}
\{g_1,g_2\}=\pm \var [\rho,g_1g_2]\, .
\eea

\section{Poisson-Lie group action on a product manifold}
 Let $\P_1$ and $\P_2$ be two Poisson manifolds 
with brackets $\{\cdot\, ,\cdot\}_{\P_1}$ and  $\{\cdot\, ,\cdot\}_{\P_2}$ that carry the Poisson action of a Poisson-Lie group $G$.
Let $\M_i:\, \P_i\to G^*$ be the corresponding non-abelian moment maps which are assumed to be Poisson.
Then, one can define the Poisson action of $G$ on the product manifold $\P=\P_1\times \P_2$ 
by taking the product\footnote{The product is naturally taken in $G^*$.} of the moment maps \cite{ASENS_1996_4_29_6_787_0}\footnote{We are grateful to  L{\'a}szl{\'o} Feh{\'e}r
for drawing our attention to this work.}
\bea
\nonumber
\M=\M_1\M_2\, , 
\eea
and allowing it to act on functions on $\P$  by means of the formula 
\bea
\xi_Xf=\langle X,\{\M,f\}_{\P}\M^{-1}\rangle\, ,~~~f\in {\rm Fun}(\P)\, ,
\eea
where $\xi_X$ is a vector field corresponding to $X\in \J$ and $\langle\cdot\, , \cdot\rangle$ is the canonical pairing between 
$\J$ and $\Jd$. We have 
\bea
\la{f_act}
\xi_Xf=\langle X,\{\M_1,f\}_{\P_1}\M_1^{-1}+\M_1\{\M_2,f\}_{\P_2}\M_2^{-1}\M_1^{-1}\rangle\, .
\eea
Let $\xi_X^{(1)}$ and $\xi_X^{(2)}$ be the fundamental vector fields induced by the group action
on $\P_1$ and $\P_2$, respectively. 
Formula (\ref{f_act}) is equivalent to the statement that at a point $x=(x_1,x_2)\in \P$, where $x_1\in \P_1$ and $x_2\in \P_2$,
the vector field $\xi_X$ is defined as 
\bea
\label{defxi}
\xi_X(x)=\xi_X^{(1)}(x_1)+\xi^{(2)}_{{\rm Ad}^*_{\M_1^{-1}(x_1)}X}(x_2)\, ,
\eea
where ${\rm Ad}^*_h$, $h\in G^*$ is the coadjoint action of $G^*$ on $G$ which is also an example of dressing transformations \cite{SemenovTianShansky:1985my}.  
One can show that the map $X\to \xi_X$, where $\xi_X$ is defined by (\ref{defxi}), is the Lie algebra homomorphism,
so that $\xi_X$ is the fundamental vector field of the group action on $G$ \cite{ASENS_1996_4_29_6_787_0,GB}. Since $G^*$ is a Poisson-Lie group,
$\M$ will have the same Poisson brackets between its entries as $\M_1$ or $\M_2$. 

To construct the Hamiltonian structure of the spin RS model, we take the product of symplectic manifolds $\P_1=D_+(G)$ and $\P_2=\Sigma_{N,\el}^{\pm}$,
\bea
\label{in_man}
\P= D_+(G) \times  \Sigma_{N,\el}^{\pm}\, .
\eea 
Here the Poisson structure on the Heisenberg double $D_+(G)$ is given by (\ref{PBAB}) and that on the oscillator manifold is \eqref{PBoscill}.
We define the Poisson action of $G$ on $\P$ through its moment map
\bea
\M\star \N=\M_+\N \M_-^{-1}\, ,
\eea
where $\N$ is the moment map \eqref{MMosc} of the action \eqref{act_ab} and $\M$ is (\ref{mmHD}). Since $\M$ and $\N$ are elements of $G^*$ modelled by $G$,
we multiply them with the star product. To obtain the RS model on the reduced phase space, we fix the moment map to the following value 
\bea
\la{Mom_map}
\M\star \N=\qq\,\mathbbm{1}\, ,
\eea
where $\mI$ is the group identity in $G$ and $\qq$ is the coupling constant. Since the right hand side of (\ref{Mom_map}) is proportional to the identity, the stability 
group of the moment map coincides with the whole group $G$ and, therefore, all the entries of $\M\star \N$ are constraints of the first class. Equation (\ref{Mom_map})
can be written as the following equation in $G$
\bea
\la{mmap}
\M=\qq\,  \omega_+ \omega_-^{-1}=\qq \, \omega\, .
\eea

Some comments are in order.
The choice of the initial manifold (\ref{in_man}), as well as the use of relevant reduction techniques to obtain the spin RS models on the reduced phase space was already
suggested earlier, see {\it e.g.} \cite{Arutyunov:1997ey, Reshetikhin:2015pma}. 
Also, a similar construction was developed in \cite{Feher:2018pmu}, where $G$ was taken to be the compact Lie group ${\rm U}(N)$. In this case
the underlying Lie bialgebra $(\J,\Jd)$ is not factorisable and the corresponding double $\double$ can be identified with the complexification of $\J=\alg{su}(N)$.
The dynamical system one finds on the reduced phase space coincides with the trigonometric spin RS model. The point, however, is that 
working with the collective spin variable $\omega$ alone leaves invisible the evolution of individual spin components of a spin vector associated to each particle. 
The aim of our present construction is to further resolve $\omega\in G^*$ in terms of internal spin degrees of freedom and obtain the dynamical equations 
for individual spins, as in \cite{Krichever:1995zw}.

\section{Reduction}
We can now develop the reduction procedure starting from the initial phase space \eqref{in_man}
\bea
\label{prod_man}
\P =D_+(G) \times \Sigma_{N,\el}^{\pm}\, .
\eea
The moment map equation (\ref{mmap}) takes the form 
\bea
\la{mmABspins}
BA^{-1}B^{-1}A=\qq(\mI+\var ab)\, .
\eea
The reduced phase space $\mathscr{P}$ is obtained by factoring solutions of (\ref{mmABspins}) by the action of the group $G$
$$
\mathscr{P}=\{{\rm Solutions~of~(\ref{mmABspins})} \}/G\, .
$$
Note that for our reduction procedure the parameter $\var$ controlling the Poisson brackets (\ref{PBAB}) of the Heisenberg double
and the brackets (\ref{PBoscill}) of the oscillator manifold is chosen to be the one and the same.

We point out that under the Poisson action on the product manifold \eqref{prod_man} the transformation of oscillators get simplified over the hypersurface defined by \eqref{mmABspins}. Indeed recalling \eqref{defxi} and \eqref{act_ab}, we get 
\begin{align}
\delta_X a_{i\alpha}\,=\, (\text{Ad}^*_{\omega \star \mathcal{M}^{-1}} X\, a)_{i\a} \qquad \delta_X b_{\alpha i}\,=\, -( b \, \text{Ad}^*_{\omega \star \mathcal{M}^{-1}} X\,)_{\a i}\, , \qquad X\in \J\,,
\end{align}
and since $\omega\star \mathcal{M}^{-1}=\omega_+ \mathcal{M}_+^{-1}\mathcal{M}_- \omega_-^{-1}\equiv \qq^{-1} \mathbbm{1}$ the action of 
${\rm Ad}^*_{\omega\star \mathcal{M}^{-1}}$ is ineffective and the oscillators transform as
\begin{align}
a_{i\alpha}\,\longrightarrow\, (h\, a)_{i\a} \qquad b_{\alpha i}\,\longrightarrow\, ( b \, h^{-1})_{\a i}\, , \qquad h=e^{X} \in G \, .
\end{align}
The most efficient way to factor out solutions by the action of $G$ is to reformulate and solve the moment map equation (\ref{mmABspins})
in terms of gauge-invariant variables. 
To this end, following \cite{Arutyunov:2019wuv} we introduce a new coordinate system on the diagonalisable locus of the Heisenberg double
\begin{align}
\la{double_decomp}
A=T\gQ T^{-1}\, , \qquad B=UP^{-1}T^{-1}\,,
\end{align}
where $\gQ$ and $P$ are diagonal matrices with entries 
\begin{align}
\gQ_{ij}\,=\, \delta_{ij}\gQ_j \qquad P_{ij}\,=\, \delta_{ij}P_j\, .
\end{align}
The matrices $T,\,U$ are Frobenius, {\it i.e.} they are subjected to the following constraints
\begin{align}
\sum_{j=1}^N T_{ij}\,=\,\sum_{j=1}^N U_{ij}=1\, ,\qquad \forall \,i=1,\dots, N\,.
\end{align}
Imposition of these constraints renders decomposition (\ref{double_decomp}) unique.

Under the transformations (\ref{act_ab}) the new variables transform as follows  
\begin{align}
\gQ~\rightarrow~ \gQ \, ,\qquad P~\rightarrow~ P\, d_T^{-1} \,d_U\, , \qquad
T~\rightarrow~ h \,T \,d_T \, , \qquad U ~\rightarrow~ h\,U\, d_U \,, 
\end{align}
where $(d_X)_{ij} = \delta_{ij}\,\sum_{k=1}^N (h X)_{ik} $ for any $X\in {\rm GL}_N(\mathbb{C})$. In particular, $\gQ$ is invariant under the $G$-action. 

Substituting  (\ref{double_decomp}) into (\ref{mmABspins}), we will get 
\bea
\nonumber
U\gQ^{-1}U^{-1}T\gQ T^{-1}=\qq(\mI+\var ab)\, ,
\eea
where, in particular, the momentum variable $P$ has completely decoupled. There are different ways to solve the above equation, we follow the one 
which relies on the simplest invariant spin variables.  We have 
\bea
\nonumber
T^{-1}U\gQ^{-1} =\qq(\gQ^{-1}T^{-1}U+\var\, T^{-1}\,  ab\, T\gQ^{-1}T^{-1}U)\, ,
\eea
Following the spinless pattern in \cite{Arutyunov:1996cmb,Arutyunov:2019wuv}, we introduce the Frobenius matrix $W=T^{-1}U$ and reintroduce the momentum $P$ by multiplying from the right both sides of the 
equations above by $P^{-1}$,
\bea
\la{12}
WP^{-1}\gQ^{-1} -\qq \gQ^{-1}WP^{-1}=\qq\var\, T^{-1}\,  ab\, A^{-1}BT\, ,
\eea
Note that under  (\ref{act_ab}) the variable $WP^{-1}$ is not invariant, rather it transforms as 
$$
WP^{-1}~\rightarrow~ d_T^{-1}(WP^{-1})d_T\, .
$$
On the other hand, a matrix $T^{-1}a$ transforms  as 
$$
T^{-1}a~\rightarrow~d_T^{-1} T^{-1}h^{-1}ha=d_T^{-1}T^{-1}a\, ,
$$ 
where we have taken into account the transformation law (\ref{act_ab}) for the spin variables. This suggests to introduce a diagonal matrix $t$ with entries
\bea
t_{ij}=\delta_{ij}\sum_{\alpha=1}^\ell (T^{-1}a)_{i\alpha}\, .
\eea
Multiplying (\ref{12}) from the left and from the right by $t^{-1}$ and $t$, respectively,  projects the moment map equation of the space of $G$-invariants 
\bea
\nonumber
t^{-1}WP^{-1}t \gQ^{-1}-\qq \gQ^{-1}\,  t^{-1}WP^{-1}t =\qq\var\,  t^{-1}T^{-1}a bA^{-1}BTt\, .
\eea
Introducing the $G$-invariant combinations
\begin{align}
\la{inv_obj}
L=t^{-1}WP^{-1}t\,Q^{-1} \, , ~\qquad \mathbf{a}=t^{-1}T^{-1}a \, , ~\qquad \mathbf{c}=bA^{-1}BTt\, ,
\end{align}
we rewrite the moment map equation in its final invariant form 
\bea
L-\qq \gQ^{-1} L\gQ=\qq\var\, \mathbf{a}\mathbf{c}\, .
\eea
The last equation is elementary solved for $L$ 
\bea
\la{L_op}
L=\qq\var \sum_{i,j=1}^N \frac{\gQ_i}{\gQ_i-\qq \gQ_j}(\mathbf{a}\mathbf{c})_{ij}\, E_{ij}\, .
\eea
The quantity (\ref{L_op}) is the Lax matrix of the hyperbolic spin RS model, as can be seen by 
by introducing the following parametrisation
\bea
\qq =e^{-2\gamma}\, ,\qquad \gQ_i=e^{2q_i}\, ,\qquad q_{ij}=q_i-q_j\, ,
\eea
so that $L$ takes the familiar form
\bea
\notag
L = \var e^{-2\gamma}\sum_{i,j}^N \frac{e^{q_{ij}+\gamma}}{2\sinh(q_{ij}+\gamma)} f_{ij} \,E_{ij}\, ,\qquad f_{ij}\equiv (\mathbf{ a} \mathbf{ c})_{ij}\,.
\eea
Computing the trace of $L^k$, 
\bea
{\rm Tr}L^k={\rm Tr} (WP^{-1}\gQ^{-1})^k={\rm Tr}(UP^{-1}T^{-1}T\gQ^{-1}T^{-1})^k={\rm Tr}(BA^{-1})^k\, ,
\eea 
we recognise that ${\rm Tr}L^k$ originate from the $G$-invariant involutive family (\ref{ham_before}). Thus, ${\rm Tr}L^k$ are in involution. We take $H=H_1$
as the Hamiltonian.

\section{Poisson brackets of $G$-invariants} 
As we have found, the reduced phase space $\mathscr{P}$ has a natural parametrisation in terms of the following $G$-invariant variables
\begin{align}
\la{ph}
\mathbf{a}_{i\alpha}\, ,\,  \mathbf{c}_{\alpha i}\, , \,  \gQ_{i}\, ,  \qquad i=1,\dots, N,\;\;\alpha=1,\dots, \el\,.
\end{align}
Note that by construction the spin variables $\mathbf{a}_{i\alpha}$ are constrained to satisfy 
\bea
\la{Frob_spin}
\sum_{\a=1}^{\el}\mathbf{a}_{i\alpha}=1\, ,
\eea
which can be regarded as the Frobenius condition in the spin space. The Lax matrix (\ref{L_op}) depends on the collective spin variables 
$f_{ij}$ only, which allows to perform the ${\rm GL}_{\ell}(\mathbb{C})$-rotations  
\bea
\nonumber
\begin{aligned}
\mathbf{a}_{i\alpha}~\rightarrow~ \frac{1}{u_i}\mathbf{a}_{i\beta}\, (g^{-1})^{\beta}_{\alpha}\, ,
\quad \mathbf{c}_{\alpha i}~\rightarrow~u_ig_{\alpha \beta}\,\mathbf{c}^{\beta}_{i}\, , \quad
u_i= \sum_{\a,\b=1}^{\ell}\mathbf{a}_{i\beta}\, (g^{-1})^{\beta}_{\alpha}\, ,
\qquad
 g\in {\rm GL}_{\ell}(\mathbb{C})\,,
\end{aligned}
\eea
without changing $f_{ij}$ and preserving the Frobenius condition (\ref{Frob_spin}).

Now we are in a position to determine the Poisson brackets between the variables  (\ref{ph}) constituting the phase space.
For that we need the Poisson brackets between $T,U,\gQ$ and $P$ variables of the double. They have been already found in our previous work \cite{Arutyunov:2019wuv}
and for the reader convenience we collect them in appendix \ref{app:double}. The brackets between invariant spins and $\gQ$ are then 
\begin{align}
\{\gQ_i,\mathbf{a}_{j \alpha}\}=0\, , \qquad \{\gQ_i,\mathbf{c}_{\alpha j}\}=\delta_{ij}\, \mathbf{c}_{\alpha j}\, \gQ_j\, .
\end{align}
For the brackets of spins between themselves we find 
\bea
\begin{aligned}
\label{ISPB}
\{\ia_1,\ia_2\}_{\pm}\,&=\, \var\big[(r^{\bullet}\mp Y) \,\ia_1\ia_2 \mp \, \ia_1\ia_2 \, {\rho} \mp \ia_1\, X_{21} \,\ia_2 \pm \ia_2 \,X_{12} \,\ia_1\big]\, , \,  \\
\{\ia_1,\ic_2\}_{\pm}\,&=\, \var\big[\ic_2(r^{\ast}_{12} \pm Y) \,\ia_1 \pm \, \ia_1 \rho_{\mp} \ic_2 \, \pm \ia_1 \ic_2\, X_{21} \mp X_{12}^{\mp} \,\ia_1 \ic_2\big] + K_{21}\,\ia_1 Z_2 - C_{12}^\rC Z_2 \,,\\
\{\ic_1,\ia_2\}_{\pm}\,&=\, \var\big[\ic_1(-r^{\ast}_{21} \pm Y) \,\ia_2 \pm \, \ia_2 \rho_{\pm} \ic_1  \mp \ia_2 \ic_1\, X_{12} \pm X_{21}^{\mp} \,\ia_2 \ic_1\big] - K_{12}\,\ia_2 Z_1 + C_{21}^\rC Z_1 \,,\\
\{\ic_1,\ic_2\}_{\pm}\,&=\, \var\big[\ic_1\ic_2\,(r^{\circ}\mp Y) \mp  {\rho} \,\ic_1\ic_2\pm  \ic_1 \,X^{\mp}_{12} \,\ic_2 \mp \ic_2\, X^{\mp}_{21}\, \ic_1\big] + \ic_2 K_{12}\, Z_1 - \ic_1 K_{21}\,Z_2\,,
\end{aligned}
\eea
where we introduced the matrices $Z=\gQ^{-1}L\gQ$ and
\bea
\begin{aligned}
X_{12}&=\sum_{i\beta\sigma\delta} (\ia_1\rho)_{i\beta\sigma\delta} \, E_{ii}\otimes E_{\sigma\delta}\, , \qquad && X^{\pm}_{12}=\sum_{i\beta\sigma\delta} (\ia_1\rho^{\pm})_{i\beta\sigma\delta}\, E_{ii}\otimes E_{\sigma\delta}\, ,\\
K_{12}&=\sum_{i \sigma} \, E_{\sigma i}\otimes E_{ii}\, , \qquad &&Y_{12}=\sum_{i\beta k \delta} (\ia_1
\ia_2\rho)_{i\beta k\delta} \, E_{ii}\otimes E_{kk}\, .
\end{aligned}
\eea
While the matrices $r^{\bullet},r^{\ast},r^{\circ}$ depend on coordinates $\gQ_i$ and they are defined as follows:
\bea
\nonumber
\begin{aligned}
r^{\bullet}&=\frac{1}{2}\sum_{i,j=1}^N \frac{\gQ_i+\gQ_j}{\gQ_i-\gQ_j} \, (E_{ii}-E_{ij})\otimes (E_{jj}-E_{ji})\, , \\
r^{\ast}&=\frac{1}{2}\sum_{i,j=1}^N \frac{\gQ_i+\gQ_j}{\gQ_i-\gQ_j} \, (E_{ij}-E_{ii})\otimes E_{jj}\, , ~~~~
r^{\circ} = \frac{1}{2}\sum_{i,j=1}^N \frac{\gQ_i+\gQ_j}{\gQ_i-\gQ_j} \, (E_{ii}\otimes E_{jj}-E_{ij}\otimes E_{ji})\, .
\end{aligned}
\eea

Writing the brackets \eqref{ISPB} for the choice ``$-$" in components one finds that for $N=1,2$ and any spin $\ell$, either $\ell=1,2$ and any number of particles $N$, it coincides with the result obtained in \cite{Chalykh:2018wce}
by means of a quasi-Hamiltonian reduction.\footnote{We thank to Maxime Fairon for pointing out the difference between the Poisson brackets \eqref{ISPB} and those of \cite{Chalykh:2018wce} for a generic choice of $N$ and $\ell$.} There are further immediate consequences of our findings. First, the rational limit of  \eqref{ISPB}, which consists in rescaling
$q_i\to \var q_i$, $\gamma\to \var\gamma$ with further sending $\var$ to zero, reproduces the Poisson structure of invariant spins established in \cite{Arutyunov:1997ey}.
Second, the Poisson algebra of collective spin variables $f_{ij}$ that follows from \eqref{ISPB}
is in general different from the result conjectured in \cite{Arutyunov:1997ey}, and their difference written in the matrix form is
\begin{align}
\label{ffY}
{\mp}\, f_{1} f_{2} \,Y \,{\mp}\,  Y f_{1} f_{2}\,  \pm\, f_{1}\, Y f_{2} \pm f_{2}\,Y f_{1}\,.
\end{align}
As a result, the Lax matrix (\ref{L_op}) does not satisfies the same Poisson algebra as in the spinless case, due to the contributions of $Y_{12}$. The Poisson bracket between Lax matrices reads
\bea
\la{LL_trig}
\frac{1}{\var}\{L_1, L_2\}_{\pm}&=& (r_{12}\mp Y) L_1 L_2- L_1 L_2 (\mr_{12} \pm Y)
+ L_1 (\bar{r}_{21}\pm Y) L_2- L_2 (\bar{r}_{12} \mp Y) L_1\,
,
\eea
where the dynamical $r$-matrices are \cite{Arutyunov:2019wuv}   
\bea
\la{r_trig}
\begin{aligned}
r &=
\sum_{i\neq j}^{\nn}\Big(\frac{\gQ_j}{\gQ_{ij}} E_{ii}-\frac{\gQ_i}{\gQ_{ij}} E_{ij}\Big)\otimes (E_{jj}-E_{ji})\, , \\
\bar{r} &=\sum_{i\neq j}^{\nn}\frac{\gQ_i}{\gQ_{ij}}(E_{ii} - E_{ij})\otimes  E_{jj}\, , 
~~~~~
\mr =\sum_{i\neq j}^{\nn}\frac{\gQ_i}{\gQ_{ij}}(E_{ij}\otimes E_{ji}-E_{ii}\otimes E_{jj})\, ,
 \end{aligned}
\eea
where similarly to the rational case we introduced the notation $\gQ_{ij}=\gQ_i-\gQ_j$.

The bracket \eqref{LL_trig} has the general form of the $r$-matrix structure compatible with involutivity of the spectral invariants of $L$, but the $\gQ$-dependent $r$-matrices of the spinless case receive now an extra contribution from the spin variables. As to the Poisson structure of \cite{Chalykh:2018wce}, the corresponding $LL$-algebra is given by \eqref{LL_trig} where $Y$ should be taken to zero.

\section{Superintegrability}
Here we explain how superintegrability of the spin RS model follows from our approach. Consider the following two families of functions on the Heisenberg double
\bea
\nonumber
J_n^{+}={\rm Tr}\big[ \S (BA^{-1})^n\big]\,   ,~~~~~  J_n^{-}={\rm Tr} \big[ \S(A^{-1}B)^n\big] \, ,~~~~n\in {\mathbb Z}\, ,
\eea
where $\S$ is an arbitrary $N\times N$-matrix which has a vanishing Poisson bracket with both $A$ and $B$.  
Using (\ref{PBAB}), it is elementary to find 
$\{H_m, J^{\pm}_n \}=0$,
where $H_m={\rm Tr}(BA^{-1})^m$ constitute a commutative family containing the Hamiltonian $H_1$. Thus, 
$J_n^{\pm}$ are integrals of motion. 
We take as $\S$ a matrix $\S^{\a\b}$ with entries $(\S^{\a\b})_{ij}=a_{i\a}b_{\b j}$. 
Thus, on the initial phase space $\P$ we have two families  of integrals 
\bea
\la{Jnpm}
J_n^{+\a\b}={\rm Tr}\big[ \S^{\a\b}(BA^{-1})^n\big]\, , ~~~  J_n^{-\a\b}={\rm Tr}\big[ \S^{\a\b}(A^{-1}B)^n\big]\, , ~~~\forall~\a,\b=1,\ldots,\el\, .
\eea
These integrals are actually functions on the reduced phase space $\mathcal{P}$ as they can be expressed in terms of gauge-invariant variables. 
Indeed, we have $BA^{-1}=Tt\,  L\, t^{-1}T^{-1}$ and $A^{-1}B=Tt(\gQ^{-1}L\gQ)t^{-1}T^{-1}$, so that 
$$
BA^{-1}=A^{-1}B(B^{-1}ABA^{-1})=A^{-1}B Tt (\gQ^{-1}L^{-1}\gQ L)t^{-1}T^{-1}
$$
and, therefore, 
\bea
\nonumber
J_n^{+\a\beta}&=&{\rm Tr}\big[ \mathbf{S}^{\a\b}\gQ^{-1}L^{-1}\gQ L^n \big]\, , ~~~~~
\nonumber
J_n^{-\a\b}=
{\rm Tr}\big[\mathbf{S}^{\a\b}\gQ^{-1}L^{n-1}\gQ\big] \, ,
\eea
where the matrix $\mathbf{S}^{\a\b}$ comprises invariant spins $( \mathbf{S}^{\a\b})_{ij}=\mathbf{a}_{i\a}\mathbf{c}_{\b j}$. 
Clearly, $J_0^{+\a\beta}=J_0^{-\a\beta}={\rm Tr}\, \S^{\a\beta}$. In the rational limit $J_n^+$ and $J_n^-$ collapse to the same conserved quantities $J_n^{\a\beta}$
introduced in \cite{Arutyunov:1997ey}.

Because $J_n^{\pm \a\b}$ are gauge invariants,
their Poisson algebra computed on $\P$ straightforwardly descends on the reduced phase space. 
To compute the Poisson brackets of the integrals,  we start with 
\bea
\nonumber
&&\frac{1}{\var}\{\S_1^{\a\b},\S_2^{\gamma\delta}\}_{\pm}=
\frac{1}{\var} C_{12}\big(\delta^{\b\gamma}\S_2^{\a\delta}-\delta^{\a\delta}\S_1^{\gamma\beta}\big)+
\rr \, \S^{\a\b}_1\S_2^{\gamma\delta} +\S^{\a\b}_1\S^{\gamma\delta}_2\, \rr - \S^{\gamma\delta}_2\, \rr_+ \S^{\a\b}_1 - \S^{\a\b}_1 \rr_- \S^{\gamma\delta}_2 \\
\label{SS}
&&~~~~\pm  \Big[ \rho_{\a\mu,\gamma \nu}\, \S_1^{\mu\beta}\S_2^{\nu\delta}+\S_1^{\a\mu}\S_2^{\gamma\nu}\rho_{\mu\beta,\nu\delta}-\S_2^{\gamma\nu}\rho_{\pm\a\mu,\nu\delta}\S_1^{\mu\beta}
-\S_1^{\a\mu}\rho_{\mp\mu\beta,\gamma\nu}\S_2^{\nu\delta}\Big]\, ,
\eea
where the indices $1,2$ are associated to the $N\times N$ matrix spaces. In deriving the last formula we used the properties 
of the spin $\rho$-matrices
$\rho^T=-\rho$ and $\rho_{\pm}^T=-\rho_{\mp}$, where $T$ means transposition. 

To present further results in a concise manner, we introduce a unifying notation 
\bea
J_n^{\a\b}={\rm Tr}(\S^{\a\b}\W^n)\, , 
\eea
where $\W$ should be identified with $\W^+=BA^{-1}$ or with $\W^-=A^{-1}B$. 
The Poisson brackets between the entries of $\W^{\pm}$ is them
\bea
\la{Wvar}
\frac{1}{\var}\{\W_1^{\pm},\W_2^{\pm}\}\,&=\, -\rr_\mp\, \W_1^{\pm} \W_2^{\pm}- \W_1^{\pm} \W_2^{\pm}\, \rr_\pm +\W_1^{\pm}\, \rr_\mp \, \W_2^{\pm}+ \W_2^{\pm} \,\rr_\pm\, \W_1^{\pm}\, .
\eea
By straightforward computation we then find the following result
\bea
\nonumber
\frac{1}{\var}\{J_n^{\a\b},J_m^{\gamma\delta}\}\, &=&\frac{1}{\var} \big(\delta^{\b\gamma}J_{n+m}^{\a\delta}-\delta^{\a\delta}J_{n+m}^{\gamma\b}\big)\\
\nonumber
&\pm &\Big[\, \rho_{\alpha \mu,\gamma\nu}J_n^{\mu\beta}J_m^{\nu\delta} +J_n^{\a\mu}J_m^{\gamma\nu}\rho_{\mu\beta,\nu\delta}  - J_m^{\gamma\nu}\rho_{\pm\alpha \mu,\nu\delta}J_n^{\mu\beta} -J_n^{\a\mu}\rho_{\mp \mu\beta, \gamma\nu}
J_m^{\nu\delta}
\Big]\\
\nonumber
&\pm&\Big[-\frac{1}{2}\big(J_n^{\a\delta}J_{m}^{\gamma\beta}-J_m^{\a\delta}J_{n}^{\gamma\beta} \big)
+ \sum_{p=0}^{m} 
\big(J_{n+m-p}^{\a\delta}J_p^{\gamma\b}-J_{m-p}^{\a\delta}J_{n+p}^{\gamma\b}\big)\Big]\, \\
\label{FR}
 &+& \frac{1\mp 1}{2}\big(J_{n+m}^{\a\delta}J_0^{\gamma\beta}-J_0^{\alpha\delta}J_{n+m}^{\gamma\beta}\big)\, .
\eea
Here the signs ``$\pm$" in the second line of this formula originate from that of (\ref{SS}) and they are associated to the choice of the 
oscillator manifold $\Sigma_{N,\ell}^{\pm}$. The different signs on the third and fourth lines have different origin and they are 
related to the choice of $\W$, namely, the upper sign corresponds to $\W^+$ and the lower one to $\W^-$. The bracket 
(\ref{FR}) is not manifestly anti-symmetric, but its anti-symmetry can be seen from the following identity
\bea
\nonumber
\sum_{p=0}^{m} 
\big(J_{n+m-p}^{\a\delta}J_p^{\gamma\b}-J_{m-p}^{\a\delta}J_{n+p}^{\gamma\b}\big)
=\sum_{p=0}^{n} 
\big(J_{n+m-p}^{\a\delta}J_p^{\gamma\b}-J_{n-p}^{\a\delta}J_{m+p}^{\gamma\b}\big)+J_{n}^{\a\delta}J_m^{\gamma\b}-J_{m}^{\a\delta}J_{n}^{\gamma\b}\, .
\eea
Further, we note that the zero modes $J_0^{\a\b}$ form a Poisson subalgebra 
\bea
\nonumber
&&\{J_0^{\a\b},J_0^{\gamma\delta}\}=\delta^{\b\gamma}J_0^{\a\delta}-\delta^{\a\delta}J_0^{\gamma\beta}\\
\nonumber
&&~~ \pm \var\Big[\, \rho_{\alpha \mu,\nu\rho}J_0^{\mu\beta}J_0^{\nu\delta} +J_0^{\a\mu}J_0^{\gamma\nu}\rho_{\mu\beta,\nu\delta}  - J_0^{\gamma\nu}\rho_{\pm\alpha \mu,\nu\delta}J_0^{\mu\beta} -J_0^{\a\mu}\rho_{\mp \mu\beta, \gamma\nu}
J_0^{\nu\delta}
\Big]\, .
\eea
Define for both choices of the sign in the last formula the quantity
\bea
\upvarpi^{\a\b}=\delta^{\a\b}+\var J_0^{\a\b}\, .
\eea
We then see that the Poisson bracket for the entries of $\upvarpi$ is nothing else but  
the Semenov-Tian-Shansky bracket in the spin space
\bea
\{\upvarpi_1,\upvarpi_2\}_{\pm}=\pm (\rho\, \upvarpi_1\upvarpi_2+\upvarpi_1\upvarpi_2\rho - \upvarpi_2\rho_{\pm}\upvarpi_1-\upvarpi_1\rho_{\mp}\upvarpi_2).
\eea
We therefore recognise that $\upvarpi$ is the non-abelian moment map for the Poisson actions (\ref{int_group})
of the spin Poisson-Lie group (\ref{spinS}) on $\Sigma_{N,\ell}^{\pm}$. Thus, $J_0^{\a\beta}$ generates infinitesimal spin transformations, 
while the conserved quantities $J_n^{\pm\a\b}$ generate higher symmetries arising from conjunction 
of spin transformations with abelian symmetries generated by $H_k$.

Since on $\mathscr{P}$ the passage from $J^{-}_n$ to $J^{+}_n$ can be understood as a redefinition of invariant spin variables, 
it is enough to consider one of these families. As is clear from (\ref{FR}), the Poisson algebra of $J^{+\a\b}_{n}$ is simpler 
because a distinguished contribution of zero modes in the last line of (\ref{FR}) decouples. Introducing a generating function of the corresponding modes
\bea
J(\lambda)=\sum_{n=0}^{\infty}J^{+}_n \lambda^{-n-1}\, ,
\eea
we then convert (\ref{FR}) into the Poisson bracket between the currents. In  
the matrix notation this bracket reads as
\bea
\la{JJ}
&&\{J_1(\lambda),J_2(\mu)\}_{\pm}\, =\frac{1}{\lambda-\mu}[C^\sC_{12},J_1(\lambda)+J_2(\mu)]\\
\nonumber
&&~~~~\pm \var \Big[\rho_{\pm}(\lambda,\mu) J_1(\lambda)J_2(\mu) +J_1(\lambda)J_2(\mu)\rho_{\mp}(\lambda,\mu)   - J_2(\mu)\rho_{\pm}J_1(\lambda) -J_1(\lambda)\rho_{\mp}
J_2(\mu)
\Big]\, .
\eea
Here we have introduce two spectral dependent $r$-matrices in the spin space 
\bea
\rho_{\pm}(\lambda,\mu)=\rho \pm \frac{1}{2}\frac{\lambda+\mu}{\lambda-\mu}C^\sC_{12} =\frac{\lambda\rho_{\pm}\mp \mu\rho_{\mp}}{\lambda-\mu}\, ,
\eea
which are the standard solutions of the trigonometric\footnote{In the difference parametrization.} Yang-Baxter equation with properties 
$$
\rho_{\pm}(\mu,\lambda)=\rho_{\mp}(\lambda,\mu)\, , ~~P\rho_{\pm}(\lambda,\mu)P=-\rho_{\pm}(\mu,\lambda)\, ,
$$
where $P=C^{{\rm s}}$ is the permutation in the spin space.
Note also that $\rho_{\pm}(\lambda,0)=\rho_{\pm}$. 

Formula (\ref{JJ}) is the symmetry algebra of non-abelian integrals of the hyperbolic spin RS model.  In the rational limit $\var\to 0$ the bracket linearises 
and coincides with the defining relations of the positive-frequency part of the ${\rm GL}(\ell)$-current algebra \cite{Arutyunov:1997ey}.  The quadratic piece of (\ref{JJ})
is the affine version of the Semenov-Tian-Shansky bracket that extends the Poisson algebra of zero modes, while the whole bracket is the Poisson pencil  of the linear and quadratic structures. 
The algebra (\ref{JJ}) has an abelian subalgebra spanned by ${\rm Tr}J(\lambda)^n$, $n\in \mathbb{Z}_+$, where the trace is taken over the spin space.

Finally, we note that the superintegrable structure of the model is ultimately responsible for the possibility to solve the equations of motion for invariant spins. Indeed, 
the equations of motion on $\P$ triggered by $H_1$ are 
$$
\dot{A}=-B\, , ~~~\dot{B}=-BA^{-1}B\, ,~~~\dot{a}=0=\dot{b}\, .
$$
These equations imply that $BA^{-1}=I$ is an integral of motion and also $a={\rm const}$, $b={\rm const}$. Thus, equations for $A$ and $B$ are elementary integrated
\bea
A(\t)=e^{-I\t}A(0)\, , ~~B(\t)=I e^{-I\t}A(0)\, .
\eea
We assume that at the initial moment of time $\uptau=0$ the system is represented by a point on the reduced phase space $\P$. In particular, at this moment of time 
coordinates of particles constitute a diagonal matrix $A(0)\equiv \gQ$ and the variables $a_{i\a}(0)\equiv a_{i\a}$ obey the Frobenius condition $\sum a_{i\a}=1$ for any $i$.
With this assumption, it is easy to see that $I=L(0)$, where $L$ is the Lax matrix containing the dependence on the initial data. 
Then, the positions of particles at time $\t$ are given by the solution $\gQ(\t)$ of the factorisation problem $e^{-L(0)\t}\gQ=T(\t)\gQ(\t)T(\t)^{-1}$, where $T(\t)$ is the 
Frobenius matrix satisfying the initial condition $T(0)=\mI$. Equations of motion for invariant spins $\mathbf{a}_{i\a}(\t)$ are then solved with the help of $T(\t)$
\bea
\nonumber
\mathbf{a}_{i\a}(\t)=\frac{T(\t)^{-1}_{ij}a_{j\a}}{\sum\limits_{\beta} T(\t)_{ij}^{-1}a_{j\beta}}=T(\t)^{-1}_{ij}a_{j\a}\, .
\eea
A similar solution can be given for invariant spins $\mathbf{c}_{\a i}$. While oscillators $a_{i\a}$ mix under the time evolution with respect to their ``particle" index $i$,
the ``spin" index $\a$ remains essentially untouched and the solution above is written for the whole  $\ell$-dimensional vector.  This situation is, of course, a consequence of the spin symmetry 
commuting with the evolution flow. 

\section*{Acknowledgements} 
We would like to thank Rob Klabbers for interesting discussions and Sylvain Lacroix for useful comments on the manuscript. G.A. is grateful to Maxime Fairon for explaining the results of \cite{Chalykh:2018wce}.  
The work of G.A. and E. O. is funded by the Deutsche Forschungsgemeinschaft (DFG, German Research 
Foundation) under Germany's Excellence Strategy -- EXC 2121 ``Quantum Universe" -- 390833306.
The work of E.O. is also supported by the DFG under the Research Training Group 1670.

\appendix
\section{Poisson structure of the Heisenberg double}
\la{app:double}
In order to compute the Poisson structure of invariant spins \eqref{ISPB}, one needs to compute the brackets on the Heisenberg double in terms of the parametrisation $(T,U,P,\gQ)$, used for the reduction. 
Indeed, recalling the expression of spins (\ref{inv_obj})
\begin{align}
\mathbf{a}=t^{-1}T^{-1}a \, , ~\qquad \mathbf{c}=bA^{-1}BTt=bT \gQ^{-1} T^{-1}UP^{-1}t\, ,
\end{align}
the needed brackets are $\{T_1,T_2\}$ for $\{\mathbf{a}_1,\mathbf{a}_2\}$ and also $\{T_1,U_2\}\, ,\,\{T_1,P_2\}\,,\,\{T_1,\gQ_2\}$ for $\{\mathbf{a}_1,\mathbf{c}_2\}$. Moreover the computation of $\{\mathbf{c}_1,\mathbf{c}_2\}$ requires the knowledge of $\{\gQ_1,\gQ_2\}$ and  $\{(A^{-1}B)_1,(A^{-1}B)_2\}$, which can be straightforwardly obtained from \eqref{PBAB}.

We introduce the following notation for $\mathcal r$-matrices (\ref{rr-mat}) dressed by generic $M,K\,\in\, {\rm GL}_N(\mathbb{C})$
\begin{align}
\mathcal r^{MK}_{\pm}\,=\,M^{-1}_1 K^{-1}_2 \,\mathcal r_{\pm}\, M_1 K_2\, ,
\end{align}
and two projectors on a generic $M_{12} \,\in\, \text{End}(\mathbb{C}^N\otimes \mathbb{C}^N)$
\begin{align}
\pi(M)_{ijkl}\,&=\, M_{ijkl}- \delta_{ij}\sum_{j=1}^N M_{ijkl}- \delta_{kl}\sum_{l=1}^N M_{ijkl} +\delta_{ij}\delta_{kl}\sum_{j,l=1}^N M_{ijkl}\, ,\\
\overline{\pi}(M)_{ijkl}\,&=\,  \delta_{kl} \,\sum_{s=1}^N\Big( M_{ijks}- \delta_{ij}\sum_{j=1}^N M_{ijks} \Big)\, .
\end{align}
Starting from \eqref{PBAB} and \eqref{double_decomp}, one can compute the required brackets $\{T_1,\gQ_2\}=0=\{\gQ_1,\gQ_2\}$ and
\begin{align}
\{T_1,T_2\}\,&=\,T_1 T_2 \,r_{\gQ}  -T_1 T_2\, \pi(\mathcal r_-^{TT} )\, ,\, \notag \\
\{T_1,U_2\}\,&= - \,T_1 U_2 \,{\pi}(\mathcal r_-^{TU})\, ,\\
\{T_1,P_2\}\,&= - \,T_1 P_2 \, \overline{r}_{\gQ} -\, T_1 P_2\,\overline \pi(\mathcal r_-^{TT})+ T_1 P_2\,\overline{\pi}(\mathcal r_-^{TU})\, , \notag 
\end{align}
where we introduced
\begin{align}
r_{\gQ}\,&=\, \sum_{i\neq j } \frac{\gQ_j}{\gQ_{ij}} \, (E_{ii}-E_{ij})\otimes (E_{jj}-E_{ji})\, ,~~~
\bar r_{\gQ}=\, \sum_{i\neq j } \frac{\gQ_j}{\gQ_{ij} } \, (E_{ii}-E_{ij})\otimes E_{jj}\, .\notag
\end{align}


\begin{thebibliography}{0}

\bibitem{Ruijsenaars:1986vq}
  S.~N.~M.~Ruijsenaars and H.~Schneider,
  ``A New Class of Integrable Systems and Its Relation to Solitons,''
  Annals Phys.\  {\bf 170} (1986) 370.
  doi:10.1016/0003-4916(86)90097-7
  
\bibitem{Ruijsenaars:1986pp}
  S.~N.~M.~Ruijsenaars,
  ``Complete Integrability of Relativistic Calogero-moser Systems and Elliptic Function Identities,''
  Commun.\ Math.\ Phys.\  {\bf 110} (1987) 191.
  doi:10.1007/BF01207363

\bibitem{Krichever:1995zw} 
  I.~Krichever and A.~Zabrodin,
 ``Spin generalization of the Ruijsenaars-Schneider model, nonAbelian 2-d Toda chain and representations of Sklyanin algebra,''
  Russ.\ Math.\ Surveys {\bf 50}, 1101 (1995)
  doi:10.1070/RM1995v050n06ABEH002632
  [hep-th/9505039].

\bibitem{Arutyunov:1997ey}
  G.~E.~Arutyunov and S.~A.~Frolov,
  ``On Hamiltonian structure of the spin Ruijsenaars-Schneider model,''
  J.\ Phys.\ A {\bf 31} (1998) 4203
  doi:10.1088/0305-4470/31/18/010
  [hep-th/9703119].

 
 \bibitem{KKS}
 D. Kazhdan, B. Kostant, and S. Sternberg, ``Hamiltonian group actions and dynamical
systems of calogero type,"  Communications on Pure and Applied Mathematics,
31 (4) 481-507, 1978.

\bibitem{Gorsky:1993dq}
  A.~Gorsky and N.~Nekrasov,
  ``Relativistic Calogero-Moser model as gauged WZW theory,''
  Nucl.\ Phys.\ B {\bf 436} (1995) 582
  doi:10.1016/0550-3213(94)00499-5
  [hep-th/9401017].
 
\bibitem{Gorsky:1994dj}
  A.~Gorsky and N.~Nekrasov,
  ``Elliptic Calogero-Moser system from two-dimensional current algebra,''
  hep-th/9401021.
  


\bibitem{Arutyunov:1996vy}
  G.~E.~Arutyunov, S.~A.~Frolov and P.~B.~Medvedev,
  ``Elliptic Ruijsenaars-Schneider model from the cotangent bundle over the two-dimensional current group,''
  J.\ Math.\ Phys.\  {\bf 38} (1997) 5682
  doi:10.1063/1.532160
  [hep-th/9608013].
  
  
\bibitem{Arutyunov:1996uw}
  G.~E.~Arutyunov, S.~A.~Frolov and P.~B.~Medvedev,
  ``Elliptic Ruijsenaars-Schneider model via the Poisson reduction of the affine Heisenberg double,''
  J.\ Phys.\ A {\bf 30} (1997) 5051
  doi:10.1088/0305-4470/30/14/016
  [hep-th/9607170].
  
\bibitem{Arutyunov:1996cmb}
  G.~E.~Arutyunov and S.~A.~Frolov,
  ``Quantum Dynamical R-matrices and Quantum Frobenius Group,''
  Commun.\ Math.\ Phys.\  {\bf 191} (1998) 15
  doi:10.1007/s002200050259
  [q-alg/9610009].


\bibitem{ACF} G.~E.~Arutyunov, L.~Chekhov and S.~Frolov,
 ``$R$-Matrix Quantization of the Elliptic Ruijsenaars-Schneider Model,"
Commun. Math. Phys., {\bf 192} (1998) 405--432.  


\bibitem{Feher:2008moa}
  L.~Feh\'{e}r and C.~Klimcik,
  ``Poisson-Lie generalization of the Kazhdan-Kostant-Sternberg reduction,''
  Lett.\ Math.\ Phys.\  {\bf 87} (2009) 125
  doi:10.1007/s11005-009-0298-3
  [arXiv:0809.1509 [math-ph]].

\bibitem{Feher:2009kp}
  L.~Feh\'{e}r and C.~Klimcik,
  ``Poisson-Lie interpretation of trigonometric Ruijsenaars duality,''
  Commun.\ Math.\ Phys.\  {\bf 301} (2011) 55
  doi:10.1007/s00220-010-1140-6
  [arXiv:0906.4198 [math-ph]].
  
\bibitem{Feher:2016nta}
  L.~Feh\'{e}r and T.~F.~G\"{o}rbe,
  ``The full phase space of a model in the Calogero-Ruijsenaars family,''
  J.\ Geom.\ Phys.\  {\bf 115} (2017) 139
  doi:10.1016/j.geomphys.2016.04.018
  [arXiv:1603.02877 [math-ph]].

\bibitem{Feher:2018pmu}
  L.~Feh\'{e}r,
  ``Poisson-Lie analogues of spin Sutherland models,''
  arXiv:1809.01529 [math-ph].


\bibitem{Chalykh:2018wce}
  O.~Chalykh and M.~Fairon,
  ``On the Hamiltonian formulation of the trigonometric spin Ruijsenaars-Schneider system,''
  arXiv:1811.08727 [math-ph].

\bibitem{Fairon:2018zgn}
  M.~Fairon,
  ``Spin versions of the complex trigonometric Ruijsenaars-Schneider model from cyclic quivers,''
  arXiv:1811.08717 [math-ph].

\bibitem{Chalykh:2017urw}
  O.~Chalykh and M.~Fairon,
  ``Multiplicative quiver varieties and generalised Ruijsenaars - Schneider models,''
  J.\ Geom.\ Phys.\  {\bf 121} (2017) 413
  doi:10.1016/j.geomphys.2017.08.006
  [arXiv:1704.05814 [math.QA]].
  
\bibitem{Arutyunov:2019wuv}
  G.~Arutyunov, R.~Klabbers and E.~Olivucci,
  ``Quantum Trace Formulae for the Integrals of the Hyperbolic Ruijsenaars-Schneider model,''
  JHEP {\bf 1905} (2019) 069
  doi:10.1007/JHEP05(2019)069
  [arXiv:1902.06755 [hep-th]].
  

\bibitem{SemenovTianShansky:1985my}
  M.~A.~Semenov-Tian-Shansky,
  ``Dressing transformations and Poisson group actions,''
  Publ.\ Res.\ Inst.\ Math.\ Sci.\ Kyoto {\bf 21} (1985) 1237.
  doi:10.2977/prims/1195178514


\bibitem{ASENS_1996_4_29_6_787_0}
H.~Flaschka and T.~Ratiu.
 ``A convexity theorem for poisson actions of compact Lie groups,''
Annales scientifiques de l'\'Ecole Normale Sup\'erieure, Ser.
  {\bf 4}  (1996) 29(6):787--809.

\bibitem{GB}
G.~Arutyunov, Elements of classical and quantum integrable models, Springer, to appear.



\bibitem{Reshetikhin:2015pma}
  N.~Reshetikhin,
  ``Degenerately Integrable Systems,''
  J.\ Math.\ Sci.\  {\bf 213} (2016) no.5,  769
  doi:10.1007/s10958-016-2738-9
  [arXiv:1509.00730 [math-ph]].


\end{thebibliography}
\end{document}